\documentclass[twocolumn,preprintnumbers,amsmath,amssymb,prb]{revtex4}

\usepackage{graphicx}
\usepackage{dcolumn}
\usepackage{bm}
\usepackage{amssymb}
\usepackage{epstopdf}
\usepackage{color}
\usepackage{dsfont}
\usepackage{float}

\begin{document}
\title{Quantum ring with the Rashba spin-orbit interaction in the regime of strong light-matter
coupling}
\author{V. K. Kozin$^{1,2}$}
\author{I. V. Iorsh$^2$}
\author{O. V. Kibis$^{3,1}$}\email{Oleg.Kibis(c)nstu.ru}
\author{I. A. Shelykh$^{1,2}$}
\affiliation{$^1$Science Institute, University of Iceland, Dunhagi
3, IS-107, Reykjavik, Iceland}\affiliation{$^2$ITMO University,
Saint Petersburg 197101, Russia}\affiliation{${^3}$Department of
Applied and Theoretical Physics, Novosibirsk State Technical
University, Karl Marx Avenue 20, Novosibirsk 630073, Russia}

\begin{abstract}
We developed the theory of electronic properties of semiconductor
quantum rings with the Rashba spin-orbit interaction irradiated by
an off-resonant high-frequency electromagnetic field (dressing
field). Within the Floquet theory of periodically driven quantum
systems, it is demonstrated that the dressing field drastically
modifies all electronic characteristics of the rings, including
spin-orbit coupling, effective electron mass and optical response.
Particularly, the present effect paves the way to controlling the
spin polarization of electrons with light in prospective
ring-shaped spintronic devices.
\end{abstract}
\maketitle

\section{Introduction}
Rapidly developing field of spintronics deals with spin related
phenomena in mesoscopic
transport~\cite{Loss,Zutic,Awschalom,Cahay,Szaszko2015}.
Generally, the spins of individual carriers can be controlled
either by application of an external magnetic field or via change
of the strength of the spin-orbit interaction (SOI) in the system.
The second approach forms the basis for so-called non-magnetic
spintronics which attracts enormous interest of the scientific
community. Particularly, the two mechanisms of the SOI are
relevant for semiconductor structures: The Dresselhaus
SOI~\cite{Dresselhaus} caused by the inversion asymmetry of the
crystal lattice and the Rashba
SOI~\cite{Rashba,BychkovRashba,BerciouxReview,Winkler} originated
from the inversion asymmetry of the structure as a whole. The
latter mechanism is of specific interest for spintronic
applications since it becomes dominant in conventionally used
InAs/GaSb-, AlSb/InAs- and GaAs/GaAlAs-based
nanostructures~\cite{Miller2003,Studenikin2003,Ghosh2004} and can
be easily tuned by an external gate voltage
\cite{Nitta1997,Engels1997,Heida1998}.
\begin{figure}[h]
\includegraphics[width=\linewidth]{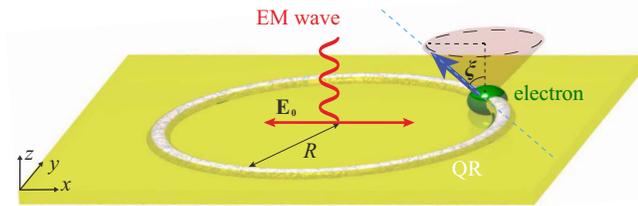}
\caption{Sketch of the system under consideration: The quantum
ring (QR) with the radius $R$ irradiated by a linearly polarized
electromagnetic (EM) wave with the electric field amplitude $E_0$.
The electron spin (the dark blue arrow) is directed along the
local quantization axis (the dashed blue line) with the spin angle
$\xi$.} \label{fig:sketch}
\end{figure}
Recently, the alternative way of tuning SOI by purely optical
methods was developed~\cite{Pervishko2015,Sheremet2016,Yudin2016}.
It is based on the regime of strong light-matter coupling when the
system ``electron + electromagnetic field'' cannot be divided into
weakly interacting optical and electronic subsystems. As a
consequence, the hybrid electron-field object --- so-called
``electron dressed by electromagnetic field'' (dressed electron)
--- appears as an elementary quasiparticle~\cite{Cohen,Scully}. The
physical properties of dressed electrons can differ sufficiently
from their ``bare'' counterparts as it has been demonstrated for
wide variety of condensed-matter structures, including bulk
semiconductors~\cite{Goreslavskii1969,Vu2004,Vu2005}, quantum
wells~\cite{Mysyrowicz,Wagner2010,Teich2013,Kibis2014,Morina2015},
quantum
rings~\cite{Kibis2011,Joibari2014,Sigurdsson2014,Sigurdsson2015,Hasan2016,Kozin2017},
graphene~\cite{Lopez2008,Oka2009,Kibis2010,Kitagawa2011,Usaj2014,Kristinsson2016,Kibis2016,Kibis2017,Iorsh2018},
topological insulators~\cite{Lindner2011}, etc. From viewpoint of
spintronic applications, it is crucially important that the SOI
strength can be modified by laser irradiation~\cite{Yudin2016}
since this allows direct optical tuning of spin relaxation time in
2D electron gas \cite{Pervishko2015} and, therefore, paves the way
to optically controlled spintronic devices~\cite{Sheremet2016}.

Although the first ferromagnetic spintronic device (the Datta-Das
spin transistor~\cite{DattaDas,Koo2009}) has been realized
experimentally, its technological production remains challenging
due to the difficulties with the efficient spin injection from
ferromagnetic contacts. Therefore, design of non-magnetic
spintronic devices which do not need the presence of ferromagnetic
elements is still an actual problem. As a possible way to solve
the problem, it was proposed to use semiconductor quantum rings
(QRs) with the Rashba SOI which induces the phase shift between
the spin waves propagating in the clockwise and counterclockwise
directions. In turn, this results in the large conductance
modulation due to the interference of the spin
waves~\cite{Bergsten2006}. As a consequence, physical basis of
various QR-based non-ferromagnetic spintronic devices
--- including spin transistors, spin filters and quantum
splitters
--- appears~\cite{Nitta1999,Nitta2003,Frustaglia2004,Shelykh2005,AronovGellerRings,PoppSpinFilter,
KisilevSpinFilter,
ShelykhSplitter,Molnar2004,Foldi2005,Foldi2006,Citro2006,Natano2007,Matityahu2013,MatityahuTarucha2013,Matityahu2017}.
In the aforecited previous studies on the subject, the spin
properties of QRs were assumed to be controlled by gate voltage.
As to the optical methods of the spin control of QRs, they are
escaped attention before. The present theoretical research is
aimed to fill partially this gap in the spintronics of QRs.

The paper is organized as follows. In Sec.~II, we derived the
effective Hamiltonian of the irradiated QR with the Rashba
spin-orbit interaction within the Floquet theory of periodically
driven quantum systems. In Sec.~III, the elaborated theory is
applied to analyze spin and optical characteristics of the
irradiated QR. As to Sec. IV, it contains conclusions and
acknowledgements.

\section{Model}
To describe an irradiated QR (see Fig.~1), we have to start from
the Hamiltonian describing an irradiated two-dimensional (2D)
electron system with the Rashba spin-orbit
interaction~\cite{Pervishko2015}
\begin{equation}\label{H2D}
\hat{\cal
H}_{\mathrm{2D}}=\frac{(\hat{\mathbf{p}}-e\mathbf{A})^2}{2m}+\alpha[\sigma_x
(\hat{p}_y-eA_y)-\sigma_y (\hat{p}_x-e A_x)],
\end{equation}
where $\hat{\mathbf{p}}=(\hat{p}_x,\hat{p}_y)$ is the operator of
electron momentum, $m$ is the effective electron mass, $e$ is the
electron charge, $\alpha$ is the Rashba spin-orbit coupling
constant, $\sigma_{x,y,z}$ are the Pauli matrices,
$\mathbf{A}=(A_x,A_y)=([E_0/\omega]{\cos\omega t},0)$ is the
vector potential of a linearly polarized electromagnetic wave
(dressing field) in the 2D plane, $E_0$ is the electric field
amplitude of the wave, and $\omega$ is the wave frequency which is
assumed to be far from resonant electron frequencies. Applying the
standard approach~\cite{CorrectHamiltonianRing} to transform the
2D electron system into the one-dimensional (1D) ring-shaped one,
we arrive from the 2D Hamiltonian (\ref{H2D}) at the Hamiltonian
of irradiated QR,
\begin{equation}\label{HQR}
\hat{\cal H}_{\mathrm{QR}}=\hat{\cal
H}^\prime+\left[\displaystyle\sum_{n=1}^{2}\hat{V}_ne^{in\omega
t}+\mathrm{H.c.}\right],
\end{equation}
where the stationary part,
\begin{equation}\label{HQR0}
\hat{\cal H}^\prime=\frac{\hat{l}_z^2}{2mR^2}
+\frac{\alpha}{R}\left[\sigma_\rho\hat{l}_z
-i\hbar\frac{\sigma_\varphi}{2}\right]+\frac{e^2E_0^2}{4m\omega^2},
\end{equation}
is the Hamiltonian of the unperturbed QR up to a field-induced
constant shift of energy,
\begin{eqnarray}\label{V1}
\hat{V}_1&=&\frac{e E_0}{2 m R
\omega}\left(\sin{\varphi}\,\hat{l}_z-i\hbar\frac{\cos{\varphi}}{2}\right)+\frac{\alpha e E_0}{2\omega}\sigma_y,\label{V1}\\
\hat{V}_2&=&\frac{e^2 E_0^2}{8 m \omega^2},\label{V2}
\end{eqnarray}
is the periodical part with the two harmonics originated from the
irradiation, $R$ is the QR radius,
$\hat{l}_z=-i\hbar\,\partial/\partial\varphi$ is the operator of
angular momentum along the $z$-axis, $\varphi$ is the polar angle
of an electron in the QR,
$\sigma_\rho=\cos{\varphi}\,\sigma_x+\sin{\varphi}\,\sigma_y$ and
$\sigma_{\varphi}=-\sin{\varphi}\,\sigma_x+\cos{\varphi}\,\sigma_y$
are the Pauli matrices written in the polar coordinates. Applying
the conventional Floquet-Magnus approach~\cite{FM1,FM2,FM3} to
renormalize the Hamiltonian of irradiated QR and restricting the
consideration by the leading terms in the high-frequency limit, we
can reduce the time-dependent Hamiltonian (\ref{HQR}) to the
effective time-independent one,
\begin{equation}\label{H}
\hat{{\cal H}}=\hat{{\cal
H}}^\prime+\sum_{n=1}^{2}\frac{\left[\hat{V}_n,\hat{V}^\dagger_{n}\right]}{\hbar
n\omega}+\sum_{n=1}^{2}\frac{ \left[\left[\hat{V}_n,\hat{\cal
H}^\prime\right],\hat{V}^\dagger_{n}\right]+\mathrm{H.c.}}{2(\hbar
n\omega)^2}.
\end{equation}
Substituting Eqs.~(\ref{HQR0})--(\ref{V2}) into Eq.~(\ref{H}), one
can rewrite the effective Hamiltonian (\ref{H}) as
\begin{equation}\label{Hef}
\hat{{\cal H}}=\hat{{\cal H}}_0+\hat{V},
\end{equation}
where
\begin{eqnarray}\label{H0}
\hat{\cal H}_0&=&\frac{\hat{l}_z^2}{2m^\ast R^2}
+\frac{\alpha}{R}\left[\sigma_\rho\hat{l}_z
-i\hbar\frac{\sigma_\varphi}{2}\right]-\left(\frac{e E_0 \alpha}{R
\omega^2}\right)^2\frac{\hat{l}_z\sigma_z}{m\hbar}\nonumber\\
&+&\frac{e^2 E_0^2}{4 m\omega^2}+\frac{1}{2m}\left(\frac{\hbar e
E_0}{4m R^2 \omega^2}\right)^2,
\end{eqnarray}
\begin{eqnarray}\label{V}
\hat{V}&=&
\left[\frac{3}{16}\gamma_1^2\cos2\varphi-\gamma_1^2\gamma_2\left(\gamma_2^2-\frac{1}{4}\right)i\sigma_x\sin\varphi\right]
\frac{\hbar^2}{2mR^2}\nonumber\\
&+&\left[\frac{i\gamma_1^2
\sin2\varphi}{2}-2\gamma_1^2\gamma_2\left(\gamma_2^2-\frac{1}{4}\right)\sigma_x\cos\varphi\right]
\frac{\hbar\,\hat{l}_z}{2mR^2}\nonumber\\
&+&\frac{\gamma_1^2\cos 2\varphi}{8mR^2}\,\hat{l}_z^2,
\end{eqnarray}
where
\begin{equation}\label{m}
m^\ast=\frac{m}{1+3\left({e E_0}/{2m R \omega^2}\right)^2}
\end{equation}
is the effective electron mass renormalized by the irradiation,
$\gamma_1= |e| E_0/(m R \omega^2)$ is the dimensionless parameter
describing the strength of electron-field coupling, and $\gamma_2=
m R \alpha/\hbar$ is the dimensionless parameter describing the
strength of Rashba spin-orbit coupling. As expected, the
Hamiltonian (\ref{Hef}) exactly coincides with the Hamiltonian of
unirradiated QR~\cite{CorrectHamiltonianRing} in the absence of
the field ($E_0=0$).

It should be noted that all effects originated from the direct
spin interaction with magnetic component of the dressing field
(particularly, the Zeeman effect and the Aharonov-Bohm effect) are
relativistically negligible since the amplitude of magnetic
induction of the field, $B_0=E_0/c$, is very small for reasonable
field intensities. Therefore, they are omitted in the developed
theory. We also neglected effects arisen from overlying electronic
modes, assuming the typical distance between transverse electronic
minibands (tens meV for state-of-the-art QRs~\cite{Lorke_2000}) to
be sufficiently large than the photon and electron energies under
consideration.

\section{RESULTS AND DISCUSSION}
To consider the Schr\"odinger problem with the effective
Hamiltonian (\ref{Hef}), let us start from its part (\ref{H0}).
Two exact eigenstates of the Hamiltonian (\ref{H0}) can be written
as
\begin{align}\label{F1}
\Psi_{1}(\varphi) = e^{i j_z\varphi}  \left(
\begin{array}{cc}
\cos(\xi/2)e^{-i\varphi/2}\\
-\sin(\xi/2)e^{i\varphi/2}
\end{array}
\right)
\end{align}
and
\begin{align}\label{F2}
\Psi_{2}(\varphi) = e^{i j_z\varphi}  \left(
\begin{array}{cc}
\sin(\xi/2)e^{-i\varphi/2}\\
\cos(\xi/2)e^{i\varphi/2}
\end{array}
\right),
\end{align}
where
\begin{equation}\label{xi}
{\xi}=\arctan\left[\frac{2 m^\ast R
\alpha/\hbar}{2({m^\ast}/{m})\left({e E_0 \alpha}/{\omega^2
\hbar}\right)^2+1}\right]
\end{equation}
is the angle between the local spin quantization axis and the
$z$-axis (see Fig.~1). It follows from single-valuedness of the
eigenstates, $\Psi_{1,2}(\varphi)=\Psi_{1,2}(\varphi+2\pi)$, that
the $z$-component of total angular momentum of electron, $j_z$,
must satisfy the condition, $j_z=\lambda n+1/2$, where
$n=0,1,2,...$ is the orbital quantum number corresponding to the
electron rotation in QR, and the sign $\lambda=\pm$ describes the
direction of the rotation (counterclockwise/cloclwise). Omitting
constant terms which only shift the zero energy, one can write the
electron energy spectrum of the eigenstates (\ref{F1})-(\ref{F2})
as
\begin{align}\label{En}
&\varepsilon^s_{\lambda n}=\frac{\hbar^2}{2m^\ast
R^2}\left(\lambda n+\frac{1}{2}\right)^2+\frac{\hbar^2}{2m^\ast
R^2}\left|\lambda
n+\frac{1}{2}\right|s\nonumber\\
&\times\sqrt{\left[2\left(\frac{m^\ast}{m}\right)\left(\frac{eE_0\alpha}{\hbar\omega^2}\right)^2+1\right]^2+\Bigg[\frac{2\alpha
m^\ast R}{\hbar}\Bigg]^2},
\end{align}
where $s=\pm1$ is the quantum number describing the spin direction
along the local quantization axis (see Fig.~1) and the spin $s=+1$
corresponds to the greater energy (\ref{En}). Within the
conventional notation~\cite{NatureRing} based on the three quantum
numbers, $|n,\lambda, s\rangle$, the eigenstates
(\ref{F1})-(\ref{F2}) can be written as
\begin{align}
&|n,+,-1\rangle = e^{i n\varphi}   \left(
\begin{array}{cc}
\cos(\xi/2)\\
-\sin(\xi/2)e^{i\varphi}
\end{array}
\right),\label{1}\\
&|n,+,+1\rangle = e^{i n\varphi} \left(
\begin{array}{cc}
\sin(\xi/2)\\
\cos(\xi/2)e^{i\varphi}
\end{array}
\right),\\
&|n,-,+1\rangle = e^{-i n\varphi}    \left(
\begin{array}{cc}
\cos(\xi/2)\\
-\sin(\xi/2)e^{i\varphi}
\end{array}
\right),\\
&|n,-,-1\rangle = e^{-i n\varphi}  \left(
\begin{array}{cc}
\sin(\xi/2)\\
\cos(\xi/2)e^{i\varphi}
\end{array}
\right),
\end{align}
for $n=1,2,3...$ and
\begin{align}
&|0,+,-1\rangle = \left(
\begin{array}{cc}
\cos(\xi/2)\\
-\sin(\xi/2)e^{i\varphi}
\end{array}
\right),\\
&|0,+,+1\rangle = \left(
\begin{array}{cc}
\sin(\xi/2)\\
\cos(\xi/2)e^{i\varphi}
\end{array}
\right)\label{2}
\end{align}
for $n=0$. It follows from Eq.~(\ref{En}), particularly, that
$\varepsilon^s_{-n}=\varepsilon^s_{n-1}$. This means that the
states $|n,-,s\rangle$ and $|n-1,+,s\rangle$ are degenerated.

The eigenstates and eigenenergies (\ref{F1})--(\ref{2}) can be
easily verified by direct substitution into the Schr\"odinger
equation with the Hamiltonian (\ref{H0}). However, the total
effective Hamiltonian (\ref{Hef}) consists of the two parts,
including both the discussed Hamiltonian $\hat{{\cal H}}_0$ and
the term $\hat{V}$. Therefore, we have to analyze the effect of
the term $\hat{V}$ on the found solutions of the Schr\"odinger
problem with the Hamiltonian $\hat{{\cal H}}_0$. It follows from
Eqs.~(\ref{V}) and (\ref{1})--(\ref{2}) that $\langle
n',\lambda',s'|\hat{V}|n,\lambda,s\rangle \sim \delta_{\lambda'
\lambda}$ for $n,n'\ge 1$. Thus, the term $\hat{V}$ does not split
the degenerate states $|n,-,s\rangle$ and $|n-1,+,s\rangle$. It
should be noted also that the discussed regime of strong
light-matter coupling is conventionally defined as a light-induced
renormalization of electronic properties without the light
absorption by electrons (see, e.g., the discussion in
Ref.~\onlinecite{Morina2015}). Particularly, the main absorption
mechanism for semiconductor structures dressed by an off-resonant
electromagnetic field
--- the collisional absorption of the field by conduction electrons
--- can be neglected if $\omega\tau\gg1$ , where $\tau$ is the electron relaxation
time~\cite{Kibis2014}. Therefore, we have to consider the case of
high-frequencies, $\omega$, when the condition $\gamma_1\ll1$ can
take place. It follows from this that the discussed term
$\hat{V}\sim\gamma_1^2$ can be considered as a weak perturbation
for a broad range of QR parameters. Particularly, the conventional
criterion of perturbation theory,
\begin{equation}\label{pert}
\left|\frac{\langle
n',\lambda',s'|\hat{V}|n,\lambda,s\rangle}{\varepsilon^{s'}_{\lambda'
n'}-\varepsilon^s_{\lambda n}}\right| \ll 1,
\end{equation}
can be satisfied for the first tens of energy levels (\ref{En}) in
the typical case of InGaAs-based QRs with the effective mass
$m=0.045m_e$, radius $R\approx200$~nm and the Rashba coupling
constant $\alpha\approx10^4$ m/s. As a consequence, the effective
Hamiltonian (\ref{Hef}) can be reduced to the simplified
Hamiltonian (\ref{H0}). Correspondingly, the found eigenstates and
eigenenergies (\ref{F1})--(\ref{2}) can be applied to describe
electronic properties of the irradiated QR.

It follows from the Hamiltonian (\ref{H0}) that the irradiation of
QR results in the two main effects: First, it renormalizes the
electron effective mass (\ref{m}) and, second, it leads to the
unusual spin-orbit coupling $\sim l_z\sigma_z$ described by the
third term of the Hamiltonian (\ref{H0}). In turn, these effects
lead to the dependencies of the spin angle (\ref{xi}) and the
energy levels (\ref{En}) on the irradiation intensity, which are
plotted in Fig.~2. It follows from Fig.~2a that the irradiation
very strongly effects on the spin angle (\ref{xi}). Namely, the
relatively weak irradiation can decrease the angle to tens
percents of its initial value in the unirradiated QR,
$\xi_0=\arctan{(2\alpha mR/\hbar)}$. Since the modulation of spin
orientation by various external actions lies in the core of modern
spintronics~\cite{Loss,Zutic,Awschalom,Cahay}, the found strong
dependence of the spin polarization on the irradiation can be
used, particularly, in prospective ring-shaped spintronic devices
operated by light. It follows from Fig.~2b that the irradiation
also strongly effects on the energy of the electron levels in QR
and their spin splitting. Such a light-induced modification of the
energy spectrum (\ref{En}) can manifest itself, particularly, in
optical measurements discussed below.
\begin{figure}[h]
\includegraphics[width=\linewidth]{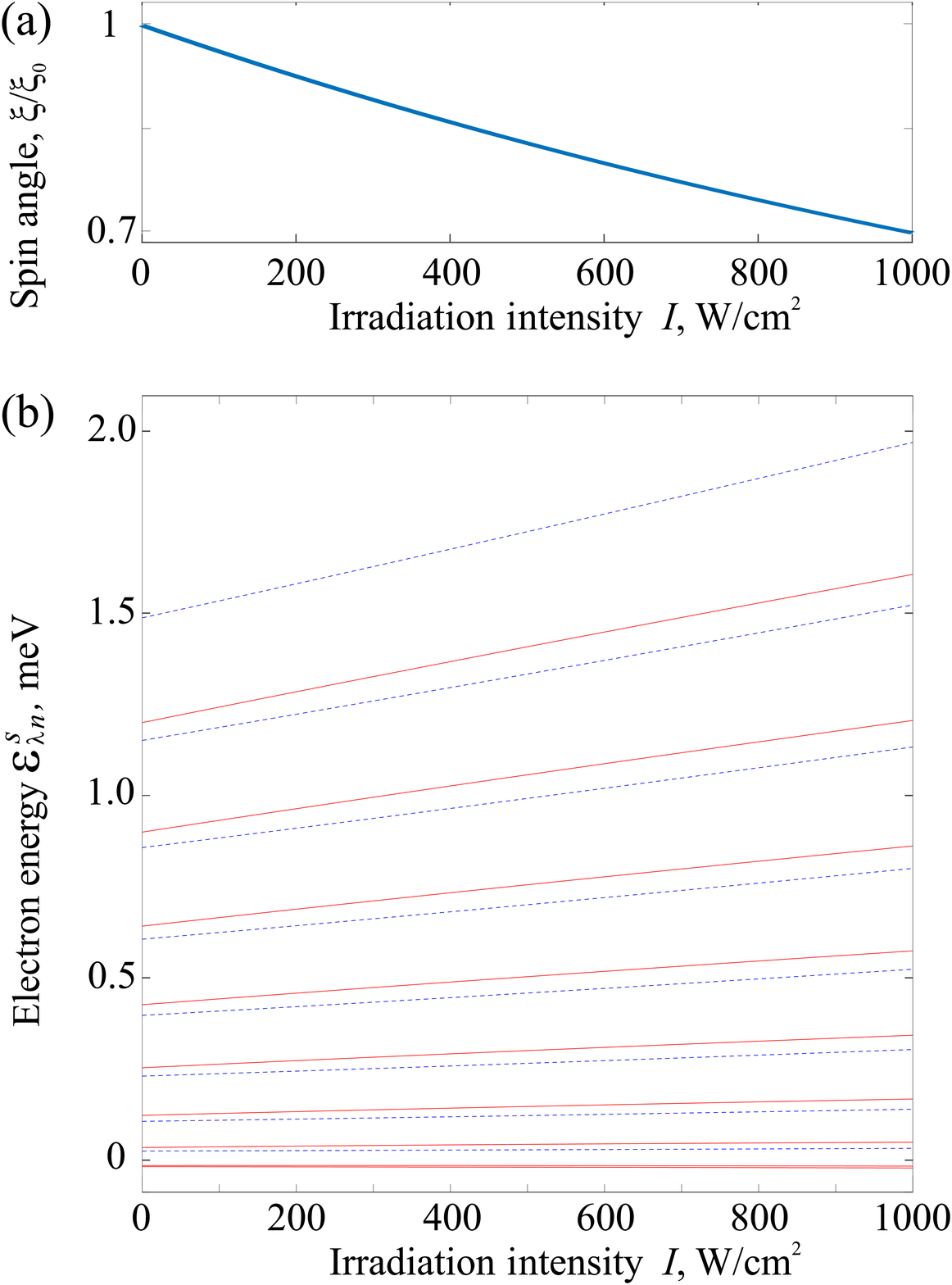}
\caption{ Electronic characteristics of InGaAs-based QR (the
electron effective mass is $m=0.045m_0$, the Rashba coupling
constant is $\alpha=10^4$~m/s and the QR radius is $R=200$~nm)
irradiated by a dressing field with the frequency
$\omega=1.6\cdot10^{12}\text{ rad/s}$: (a) Dependence of the spin
angle, $\xi$, on the irradiation intensity, $I$; (b) Dependence of
the first nine electron energy levels, $\varepsilon^s_{\lambda
n}$, on the irradiation intensity, $I$, for the counterclockwise
electron rotation in the ring ($\lambda=+$), where the dashed and
solid lines correspond to the different spin orientations
($s=\pm1$).} \label{fig:energy_levels}
\end{figure}

Let us consider a QR irradiated by a two-mode electromagnetic wave
consisting of a strong dressing field (which renormalizes the
energy spectrum of electrons according to the aforesaid) and a
relatively weak probe field with the frequency $\Omega$ (which can
detect the discussed renormalization of the energy spectrum). The
optical spectrum of absorption of the probe field can be obtained
with using the conventional Kubo formalism~\cite{Kubo}. Within
this approach, the longitudinal conductivity describing the
response of the QR to the probe field polarized along the $j=x,y$
axis reads
\begin{eqnarray}\label{Kubo}
\sigma_{jj}&=&\sum_{\substack{n,\lambda,s\\n',\lambda',s'}}\frac{(f(\varepsilon^{s'}_{\lambda'
n'})-f(\varepsilon^s_{\lambda n}))|\langle n',\lambda',s'|
\hat{v}_j|n,\lambda, s\rangle|^2}{(\varepsilon^{s'}_{\lambda'
n'}-\varepsilon^s_{\lambda n})(\varepsilon^{s'}_{\lambda'
n'}-\varepsilon^s_{\lambda n}+\hbar\Omega+i\Gamma)}\nonumber\\
&\times&\frac{\hbar e^2}{i \pi R^2},
\end{eqnarray}
where $f(\varepsilon)$ is the Fermi-Dirac distribution function,
$\hat{v}_j=\hat{p}_j/m$ is the velocity operator, and
$\Gamma=\hbar/\tau$ is the broadening of energy levels depending
on the electron relaxation time, $\tau$. It should be noted that
the used spin index, $s=\pm1$, describes the spin projection on
the local quantization axis (see the dashed line in Fig.~1) which
depends on the electron location in QR and, correspondingly,
depends on the direction of the vector of electron velocity. As a
consequence, the matrix of the velocity operator in
Eq.~(\ref{Kubo}), $\langle
n^{\prime},\lambda^{\prime},s^{\prime}\lvert\hat{v}_j\rvert
n,\lambda,s\rangle$, is not diagonal in this spin index.
Particularly, direct calculation results in $\langle
n^{\prime},\pm,s^{\prime}\lvert\hat{v}_j\rvert
n,\pm,s\rangle\sim(\delta_{n-n^\prime,1}+\delta_{n-n^\prime,-1})$
and $\langle n^{\prime},\mp,s^{\prime}\lvert\hat{v}_j\rvert
n,\pm,s\rangle\sim\delta_{n+n^\prime,1}$. As a consequence, the
probe field can induce electron transitions between the electron
states with mutually opposite local spin directions. Substituting
Eqs.~(\ref{En})--(\ref{2}) into Eq.~(\ref{Kubo}), one can
calculate the sought absorption spectrum of the probe field (see
Fig.~3), which is represented by the real part of the
conductivity, $\mathrm{Re}(\sigma_{jj})$. In the absence of the
dressing field, the absorption spectrum of the QR plotted in
Fig.~3a consists of the three peaks corresponding to the following
electron transitions: $|5,+,+1\rangle\rightarrow|4,+,+1\rangle$,
$|7,+,-1\rangle\rightarrow|6,+,+1\rangle$ and
$|7,+,-1\rangle\rightarrow|6,+,-1\rangle$ (peak $1$);
$|6,+,+1\rangle\rightarrow|5,+,+1\rangle$,
$|8,+,-1\rangle\rightarrow|7,+,1\rangle$ and
$|8,+,-1\rangle\rightarrow|7,+,-1\rangle$ (peak $2$);
$|7,+,+1\rangle\rightarrow|6,+,+1\rangle$ and
$|9,+,-1\rangle\rightarrow|8,+,-1\rangle$ (peak $3$).
\begin{figure}[H]
\includegraphics[width=\linewidth]{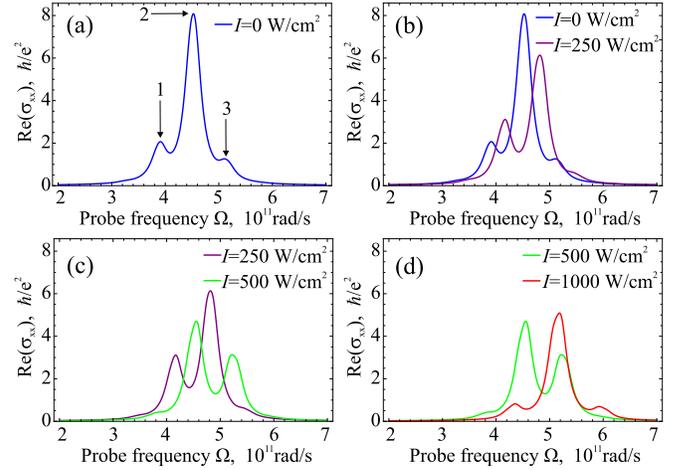}
\caption{Absorption spectra of the probe field with the frequency
$\Omega$ for the InGaAs-based QR (the electron effective mass is
$m=0.045m_0$, the Rashba coupling constant is $\alpha=10^4$~m/s,
the electron relaxation time is $\tau=70$~ps, the temperature is
$T=5$~K, the Fermi energy is $\mu=1$~meV and the QR radius is
$R=200$~nm) irradiated by a dressing field with the frequency
$\omega=1.6\cdot10^{12}\text{ rad/s}$ and different irradiation
intensities, $I$.} \label{fig:abs_spectra}
\end{figure}

The evolution of this spectrum under influence of the dressing
field is presented in Figs.~3b--3d. In the absence of the dressing
field, the highest peak $3$ originates from the transitions
$|6,+,+1\rangle\rightarrow|5,+,+1\rangle$ and
$|8,+,-1\rangle\rightarrow|7,+,-1\rangle$ since the chosen Fermi
energy, $\mu=1$ meV, lies in the middle between the corresponding
levels (see Fig.~2b). Since the dressing field increases the
distance between the energy levels (\ref{En}), it shifts the peaks
to the right and deform them (see Figs.~3b--3d). It should be
noted that the shape of the spectrum at the irradiation intensity
$I=1000$~W/cm$^2$ is very similar to case of unirradiated QR
(compare Figs.~3a and 3d). Physically, the similarity appears
since the Fermi energy, $\mu=1$ meV, lies at this intensity again
in the middle between the corresponding levels (see Fig.~2b).
However, the highest peak in this case arises from the peak $1$ in
Fig.~3a and, therefore, corresponds to the transitions
$|5,+,+1\rangle\rightarrow|4,+,+1\rangle$,
$|7,+,-1\rangle\rightarrow|6,+,+1\rangle$ and
$|7,+,-1\rangle\rightarrow|6,+,-1\rangle$. Finalizing the
discussion, let us formulate how the dressing field parameters
should be chosen in experiments. It follows from the Hamiltonian
(\ref{H0}) that the absolute value of the light-induced
renormalization of all electronic characteristics is proportional
to the squared ratio $E_0/\omega^2$. Therefore, we have to keep
this ratio not too small to observe the discussed renormalization
experimentally for reasonable dressing field amplitudes, $E_0$.
Particularly, this is why the dressing field frequency, $\omega$,
in Figs.~2 and 3 is chosen to be in the THz range.

\section{Conclusions}
In conclusion, we demonstrated that the key electronic
characteristics of QRs with the Rashba spin-orbit interaction ---
the structure of electron energy levels and the spin polarization
of electrons
--- strongly depend on an off-resonant irradiation. Particularly,
the modification of both electron effective mass and the
spin-orbit coupling appear. It is shown that the
irradiation-induced renormalization of electron energy spectrum
can be observed in state-of-the-art optical experiments, whereas
the light sensitivity of the spin orientation can be exploited in
prospective spintronic devices operated by light.

\begin{acknowledgements}
The work was partially supported by RISE Program (project CoExAN),
Russian Foundation for Basic Research (project 17-02-00053),
Rannis project 163082-051, Ministry of Education and Science of
Russian Federation (projects 3.4573.2017/6.7, 3.2614.2017/4.6,
3.1365.2017/4.6, 3.8884.2017/8.9, 14.Y26.31.0015) and Government
of Russian Federation (Grant 08-08).
\end{acknowledgements}

\end{document}